\documentclass[sn-mathphys-num]{sn-jnl}

\usepackage{graphicx}%
\usepackage{multirow}%
\usepackage{amsmath,amssymb,amsfonts}%
\usepackage{amsthm}%
\usepackage{mathrsfs}%
\usepackage[title]{appendix}%
\usepackage{xcolor}%
\usepackage{textcomp}%
\usepackage{manyfoot}%
\usepackage{booktabs}%
\usepackage{algorithm}%
\usepackage{algorithmicx}%
\usepackage{algpseudocode}%
\usepackage{listings}%

\theoremstyle{thmstyleone}%
\newtheorem{theorem}{Theorem}
\newtheorem{proposition}[theorem]{Proposition}%

\theoremstyle{thmstyletwo}%
\newtheorem{example}{Example}%
\newtheorem{remark}{Remark}%

\theoremstyle{thmstylethree}%
\newtheorem{definition}{Definition}%

\begin{document}

\title[A short introduction to Neural Networks and their application to Earth and Materials Science Science]{A short introduction to Neural Networks and their application to Earth and Materials Science}


\author*[1]{\fnm{Duccio} \sur{Fanelli}}\email{duccio.fanelli@unifi.it}

\author[2]{\fnm{Luca} \sur{Bindi}}\email{luca.bindi@unifi.it}

\author[1]{\fnm{Lorenzo} \sur{Chicchi}}\email{lorenzo.chicchi@unifi.it}

\author[1]{\fnm{Claudio} \sur{Pereti}}\email{claudio.pereti96@gmail.com}

\author[3]{\fnm{Roberta} \sur{Sessoli}}\email{roberta.sessoli@unifi.it}

\author[2]{\fnm{Simone} \sur{Tommasini}}\email{simone.tommasini@unifi.it}

\affil*[1]{\orgdiv{Department of Physics and Astronomy}, \orgname{University of Florence}, \orgaddress{\street{via G. Sansone 1}, \city{Sesto Fiorentino}, \postcode{50019}, \state{Italy}}}

\affil*[2]{\orgdiv{Department of Earth Sciences}, \orgname{University of Florence}, \orgaddress{\street{Via La Pira 4}, \city{Firenze}, \postcode{50121}, \state{Italy}}}

\affil*[3]{\orgdiv{Department of Chemistry U. Schiff-DICUS and INSTM Research Unit} \orgname{University of Florence}, \orgaddress{\street{via della Lastruccia 3-13}, \city{Sesto Fiorentino}, \postcode{50019}, \state{Italy}}}


\abstract{Neural networks are gaining widespread relevance for their versatility, holding the promise to yield a significant methodological shift in different domain of applied research. Here, we provide a simple pedagogical account of the basic functioning of a feedforward neural network. Then we move forward to reviewing two recent applications of machine learning to Earth and Materials Science. We will in particular begin by discussing a neural network based  geothermobarometer, which returns reliable predictions of the pressure/temperature conditions of magma storage. Further, we will turn to illustrate how machine learning tools, tested on the list of minerals from the International Mineralogical Association, can help in the search for novel superconducting materials.
}




\maketitle

\section{Introduction}\label{sec1}

Machine learning (ML) technologies \cite{bishop2006pattern, shalev2014understanding} are becoming increasingly popular due to their inherent degree of transversal adaptability, which transcends different realms of applications \cite{esteva2019guide}. Among the most
employed tools, Feedforward Neural Networks (FNNs) constitute the basic functional units for a plethora of ML implementations, spanning across hierarchic levels of complexity \cite{bishop2023deep, prince2023understanding, goodfellow2016deep}. FNNs seek at solving an optimization problem, via minimizing a suitably defined loss function which confronts the expected target to the output produced at the exit layer. In doing so, FNNs learn from data by identifying distinctive features that yield intertwisted correlations, in a context where the underlying generative model is a priori unknown. The output produced at the exit layer results from a a nested sequence of linear (across layers) and non-linear (localized on the nodes) manipulations of the data provided as an entry. The targets of the optimization are the weights of the links that connect pair of nodes belonging to adjacent layers of the deep arrangement. The set of trained weights hides inside the ability of the network to efficiently cope with the assigned task. It is indeed in the architecture of the weights, and in the ensuing activation/inhibition patterns which orchestrate its intimate functioning, that  is ultimately stored the processing ability of the FNN, a computational device often referred to as a bizarre black box. The next section is entirely devoted to dampen this veneer of mystery by providing a synthetic account of the foundational aspects that underlie the operation of a basic FFN.

FNN bears an unprecedented – still largely unexplored – potential to promote innovation in distinct domains of fundamental and applied research. To convincingly elaborate along these lines, we will here review two recent studies, relevant for Earth Sciences in general. We shall in particular begin discussion a novel Machine Learning approach (termed GAIA, from Geo Artificial Intelligence thermobArometry) to estimate P-T conditions of magma storage and migration within the crust. The developed Feedforward Neural Network scheme applied to clinopyroxene composition, a group of important rock-forming minerals, sets the basis for a novel generation of reliable geothermobarometers, which extends beyond the paradigm associated to crystal-liquid equilibrium. 
Then we will move forward to discussing a computer assisted framework to supervised classification and regression of superconductive (SC) materials which builds on a sophisticated algorithm that combines an ensemble made of interlaced FFN. By providing the chemical constituents of the examined compounds as an input to the algorithm, 
one can anticipate its potential superconducting nature and determine the associated critical temperature. The method was then applied to scan the whole list of  minerals from the International Mineralogical Association, a computer aided search that led to the discovery of the 
first certified superconducting material identified by artificial intelligence methodologies.

\section{A short introduction to Neural Networks}\label{sec2}

Machine learning (ML) defines a branch of artificial intelligence that applies algorithms to systematize  underlying relationships among data, in ways that prove useful to solve a large gallery of problems ranging from big data analytics, behavioral pattern recognition and information evolution. 

According to Tom M. Mitchell’s definition of ML \cite{TomMitchell}: “A computer program is said to learn from experience E with respect to some class of tasks T and performance measure P, if its performance at tasks in T, as measured by P, improves with experience E.” This latter quote follows the pioneering approach by Alan Turing in his seminal paper \cite{Turing1950}, where the  benchmark standard for gauging machine intelligence from an operational standpoint was {\it de facto} introduced.
 
Experience E represents therefore a key step and it is structured
following mainly three different learning paradigms: 

\begin{itemize}
\item {\bf Supervised Learning}. Supervised learning algorithms
 require a dataset made up of pairs (also called input-output pairs): instances, on the one side, and related labels, on the other. It
is customary to refer to supervised learning because, during training stages, the machine is supplied with both the input data and the desired outputs. Training consists in reducing the discrepancy between predicted and expected outputs. By implementing this strategy, the machine self-consistently extrapolates hidden relationships that link pairs of provided data and use the acquired knowledge to handle items that were not seen during training phase. 
\item {\bf Unsupervised Learning}. At variance with the above, 
unsupervised learning algorithm do not use labeled data. The algorithm must be therefore capable of autonomously extracting patterns from the supplied data. Notable tasks in which unsupervised learning proves useful 
are clustering, patterns explanation and dimensionality reduction.
\item {\bf Reinforced Learning}. The algorithm interacts
with the environment to achieve a given objective. During this
process, the machine learns to complete the assigned task 
while subject to positive (rewards) or negative (punishments) stimuli.
\end{itemize}

\begin{figure}[ht]
	\centering
	\includegraphics[width = 1\textwidth]{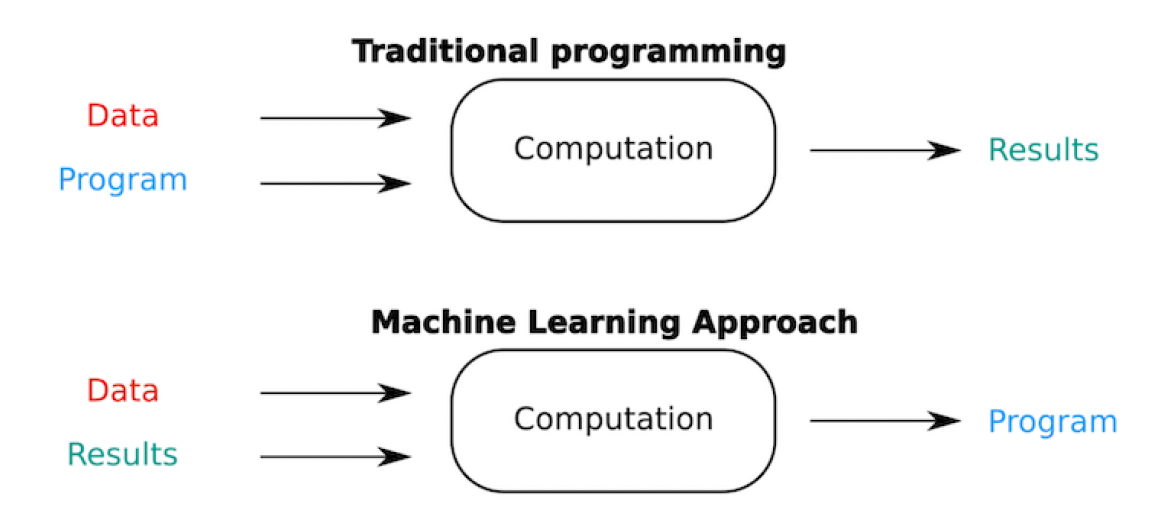}
	\caption{\footnotesize Graphical comparison of the traditional approach to programming, on the one side, and Machine Learning, on the other. According to the classical viewpoint, the programmer
instructs the machine to transform  data supplied as an input
into the desired output. In Machine Learning, the machine automatically
extrapolates the hidden rules that link data to produced results. In the first case, the output of the computation resides in the answer, while in the second it is the governing rule, drawn from a proper hypothesis space.}
\label{programming}
\end{figure}

In the following, we will make explicit reference to the case of Supervised Learning, the paradigm that has been effectively employed to carry out the analysis that we shall hereafter review. The typical Machine Learning approach, particularly in its supervised modality, overturns the classic programming conception. Canonically, the programmer's role is to provide information to the machine. This latter should operate a certain function (or set of functions/manipulations) on a given input 
and eventually return the desired result. According to the traditional approach, the operator provides therefore the machine with the input data and the necessary information to apply the sought transformation. Under the  supervised learning protocol, the programmer hands in both the input data and the desired output of the computation. It is the device, properly trained on the supplied information, that seek to identify the missing function, by effectively linking the provided data to the expected output. This is achieved by searching within the so called {\it Hypothesis space}, the set of possible functions that can be hypothetically invoked for solving the assigned task. A pictorial representation of the two alternative frameworks to which we made reference above is given in Fig. \ref{programming}.

As we will make clear in the following, algorithms can be operated according to two distinct modalities, depending on the task that we aim at solving. If the output takes the form of a discrete variable, we deal with a classification problem. The idea is to look for a hypersurface that appropriately divides the classes into which the dataset is inherently organized. If the produced output is a continuous quantity, the device is handling a regression problem, and the task of the trained algorithm is to predict the specific value associated to the instance under scrutiny. 

We will here focus on a specific class of artificial intelligence tools that implements the above prescriptions. These are the so called neural networks, dedicated algorithm inspired, to some extent, to the actual functioning of the human brain. The sought, a priori unknown function that links supplied data and expected output, is in fact encoded in the intricate web of mutually entangled units, the artificial analogues of the neurons in biological brains.

\begin{figure}
    \centering
    \includegraphics[scale=0.4]{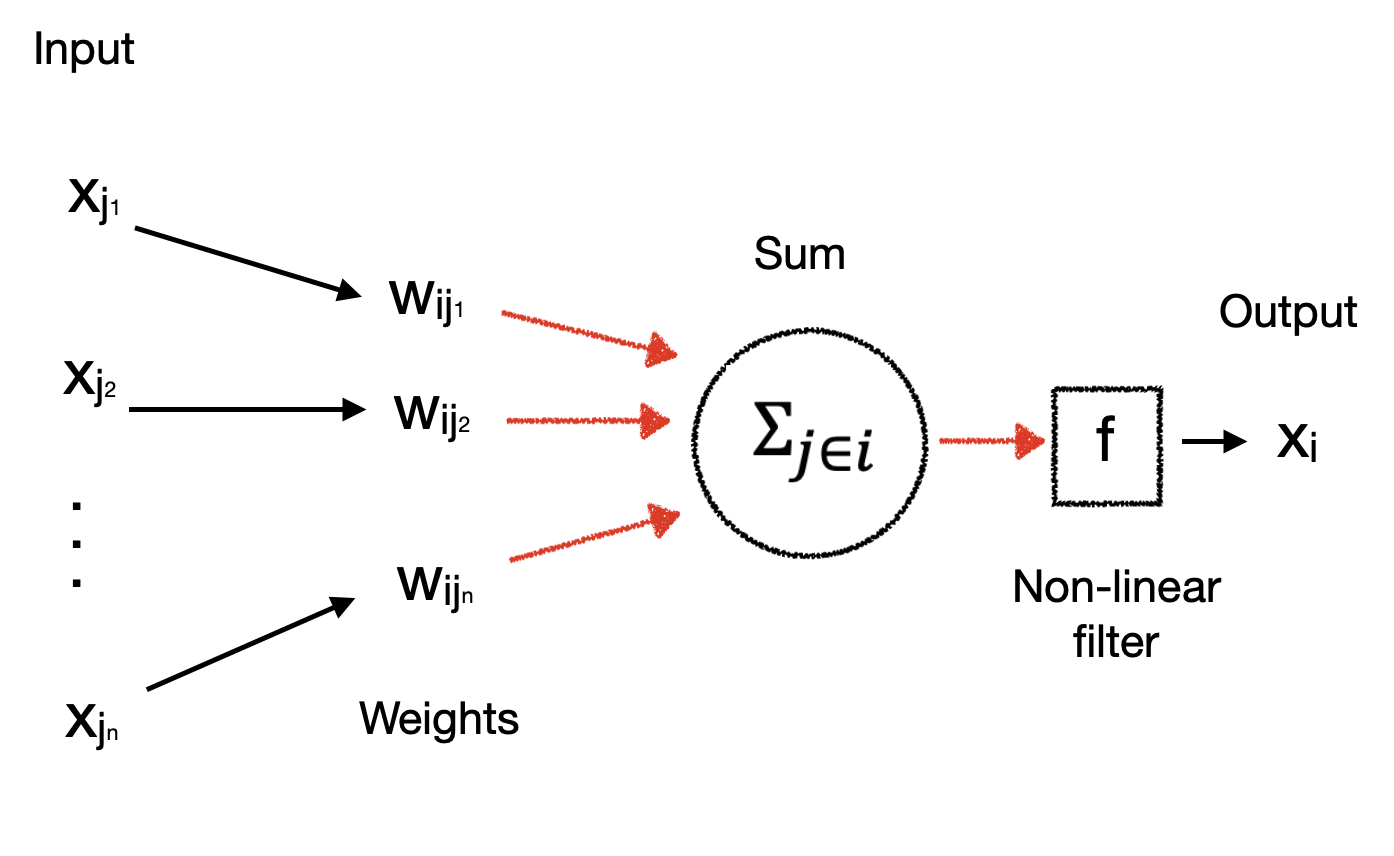}
    \caption{Schematic functioning of an artificial neuron. Input $x_{j_1}, x_{j_1}, ..., x_{j_n}$ from the $n$ neurons connected to neuron $i$ are first modulated according to the associated synaptic (positive/negative) strength $w_{i j_k}$, and then summed together (linear transformation). The result of the linear transformation defines the argument of a non linear filter (activation function) that sets the status of the receiving node $i$. If the aggregated and weighted signal crosses a given threshold, neuron $i$ fires, yielding a successive cascade of reactions. Otherwise neuron $i$ keeps silent and the information transfer is halted.}
    \label{fig:one_neuoron}
\end{figure}

Synthetic neurons, also termed nodes or computational units, can either receive input from an external source or from other neurons. As for their biological analogues, artificial neurons react when the integrated stimuli passed by their immediate neighbors cross a given threshold. To mathematically account for this scheme of interaction, we label with $x_i$ the activity associated to node (synthetic neuron) $i$ and identify with $w_{ij}$ the weight of the connection (artificial synapses) that links node $j$ to node $i$. Note that $w_{ij}$ can in principle take positive or negative values depending on the specific nature of the inter-node interaction (excitation versus inhibition). Then we stipulate:

\begin{equation}
\label{eq_neuron}
    x_i = f(\sum_i w_{ij} x_j)
\end{equation}

where $f(\cdot)$ is a suitable non linear function (chosen from a  vast arsenal of distinct possibilities, as we will clarify in the following) which is meant to mimic the sigmoidal activation profile, as displayed by real biological neurons: in the simplest possible scenario, in fact, if the aggregated signal passed by neurons $j$ (and weighted by the associated entries $w_{ij}$) is below an imposed threshold (set by the chosen function or modulated by apposite bias), the receiver $i$ stays silent. Otherwise, a response is triggered and delivered to the neurons in 
cascade with $i$, following a similar line of reasoning.  A pictorial representation of the scheme encoded via eq. (\ref{eq_neuron}) is provided in Fig. \ref{fig:one_neuoron}.

\begin{figure}
    \centering
    \includegraphics[scale=0.4]{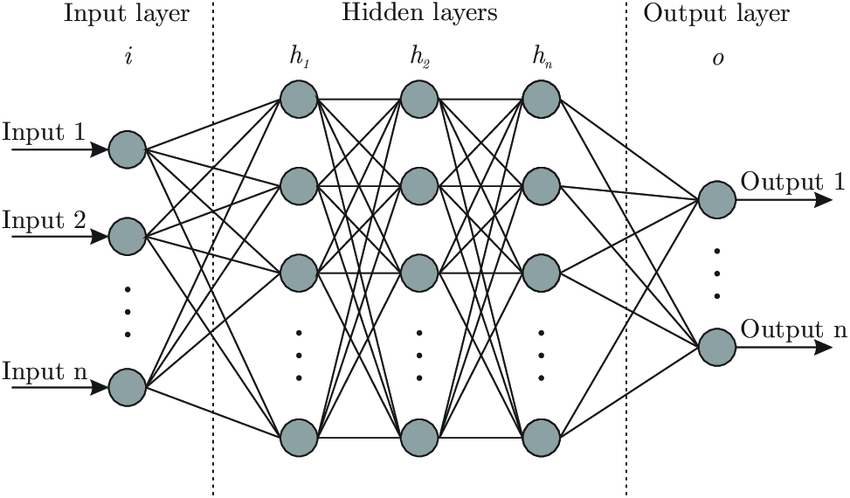}
    \caption{Pictorial representation of a neural network of the type described in the main text. Different instances (the data to be analyzed) are supplied to the \textit{Input Layer}. These are then processed by the \textit{Hidden Layers}, before eventually reaching the so called \textit{Output Layer}, the actual location where decision are taken.}
    \label{fig:my_label}
\end{figure}

The jargon used to describe artificial neural networks is variegated. Terms like artificial neural networks, deep feedforward networks, feedforward neural networks and multilayer perceptrons (MLP) are used interchangeably to point to the same mathematical object. Here, we will solely refer to feedforward neural networks, i.e. networks in which the information flows, layer after layer, in the forward direction. More concretely, a feedforward network is an ordered sequence of layers, each made of a finite collection of functional computing units, the neurons as defined above. Each neuron, from any given layer, is connected (via the aforementioned artificial synapses) to all the neurons belonging to the immediate (forward) neighbor layer. For this reason this architecture is often said to be dense (in terms of associated links) or fully connected. The activity produced as an output of a given layer, as follows the mathematical mechanism that we have discussed above with reference to just one node of the target ensemble, is sent as an input to the nodes of the next layer, see Fig. \ref{fig:my_label}. It is in this nested sequence of transformations, from the input towards the output, and specifically in the signed weights that define the synaptical web of interlinked connections, that resides the inherent ability of the (trained) neural network to cope with the assigned tasks. In short:

\begin{itemize}
    \item Input Layer: This is the layer where the input (belonging to the instances space) is supplied. The size of the input layer, say $N_0$, is bound to the dimension of the data that are to be eventually inspected. If the supplied data are images, $N_0$ stands for the number of associated pixels. Input nodes are usually assigned a linear activation function ($f$), meaning that they just mirror the entry information  to be subsequently processed. 
    \item Hidden Layers: Sequence of $k$ intermediate layers, respectively made of $N_2$, $N_3$,…, $N_k$ neurons. This is where the intermediate processing or computation take place.  Data are manipulated to extract relevant features and the processed information transferred across the network, stack by stack, following the same mathematical recipe introduced above, that alternates linear and non linear transformations. One can freely adjust the number of elements (neurons) pertaining to each stack, and/or modulate the number of hidden layers that define the architecture of the model to be trained. This amounts to shape the hypothesis space that the algorithm explores when searching the optimal score function. 
    \item Output Layer: This is the layer that returns the predictions relative to the examined instances. The number of neurons $N_{k+1}$ and the selected non linear filter depends on the specific problem being inspected. If the aim is to distribute the instances in just two classes (feedforward neural network run in classification mode), the operator should set $N_{k+1}=2$ and choose a softmax activation function, a dedicated  mathematical function that converts a vector made of real numbers into a probability distribution (stated differently, the element of the vector displayed at the exit layer are transformed so as to reflect the probability of belonging to the associated class).
\end{itemize}

\begin{figure}[!h]
    \centering
    \includegraphics[scale = 0.4]{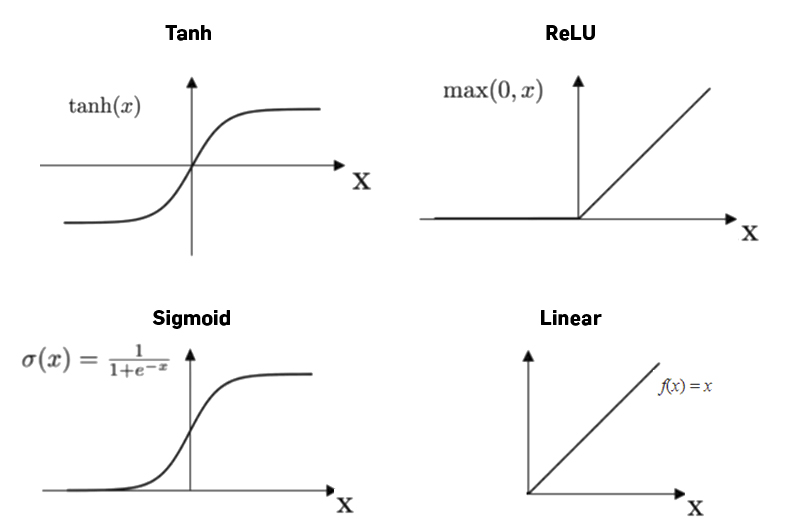}
    \caption{The most used activation functions are depicted. The horizontal axis refers to the argument of the applied non linear filter, namely to the weighted sum of signals that reaches the  target node. An additional set of scalar parameters, called bias, can be in principle introduced which yield a rigid shift in the computed activity. This supplementary effect is here neglected for the sake of simplicity. In the vertical axis the output produced by the neuron is displayed. This output will represent the input for the neurons that}
 \label{fig:funzioniattivazione}
\end{figure}

We emphasize again that the neural network is the result of two operations, a linear sum that piles up input signals (reaching every neurons of a given stack and being transmitted from all the neurons composing the preceding layer) and a non linear filter applied at the nodes location. The most popular non linear  activation functions are depicted in Fig. \ref{fig:funzioniattivazione} as a reference. The ability of the neural networks to operate as seemingly intelligent entities resides in the collection of $N_0 \times N_1 +  N_1 \times N_2 + N_2 \times N_3 +…+ N_k \times N_{k+1}$ positive and negative scalars that are to be provided for linking the supplied input to the produced input, following sequential applications of the algorithmic recipe made explicit by equation (\ref{eq_neuron}).

The next step in the story is to discuss how the weights can be chosen so as to implement the sought functionalities. The key idea is to estimate the weights in such a way that the error at the exit layer gets minimized. 
The overall procedure is made up of two independent steps. 

During the forward step, the provided input is propagated through the layers, up to the output layer where decision is to be taken. The produced output is confronted with the expected output, for a number of annotated instances that have been a priori labelled. These latter define the ground truth, namely the information provided by empirical evidence that is eventually used to fuel the inference algorithm. The distance between expected and obtained predictions, for a given set of weights (initially set at random), is encoded in the so called loss function, sometimes referred to as to the cost function.  Although many loss functions exist, all of them penalize on the distance between the predicted score (be it operated in regression or classification modalities) from a given input as confronted to its correct output, stored in the annotated dataset. Among the most popular loss function, we mention the mean squared error (MSE), the average squared difference between the estimated values (which inherently depend on the set of assigned weights that propagate forward the signal) and the actual reference. The second step of the computation, the backpropagation stage, quantifies the gradient of the loss function, with respect to the weights to be eventually adjusted, for a single input–output pair. To this end, the gradient is computed  
one layer at a time, by iterating backward from the last to the first, and following the prescriptions of the chain rule. Strictly speaking, backpropagation identifies an efficient algorithm for computing the gradient. With some abuse of language, it is however used to point to the entire learning algorithm, including how the gradient is used, such as via stochastic gradient descent, or as an intermediate step in a more elaborated optimizer strategy, such as the renewed Adam scheme \cite{kingma2017adam}. The actual training phase amounts to modulate the weights by a factor that scales proportionally to the computed gradient. The proportionality factor is the learning rate which can be empirically adjusted depending on the problem at hand. In pictorial terms, the whole process can be represented as the attempt to reach the valley of a mountain of unknown topography, starting from the summit.  A possibility is to look in every direction to see which one steeps downward the most, and then move forward in that direction (i.e. along the gradient descent). The evaluation needs to be repeated after each move to sample the local geometry of the ground, thus yielding an iterative scheme that can hopefully guide one through the bottom of the mountain. Under this simplified description, the mountain stands for the loss function and the goal is to reach the bottom of the latter by implementing a recursive gradient descend march. At this point, one recovers a self-consistent estimate of the weights which are to be assigned to the feedforward neural network for minimizing the discrepancy between expected and predicted outputs. Dedicated strategies can be devised to avoid getting stuck in a small trough or valley, and head toward the desired global minimum. Also different mathematical parametrizations of the message passing information can be possibly considered which would mold the underlying landscape yielding non trivial reflex in the numerical handling of the optimization problem. As a relevant example, we mention the spectral approach to neural network learning \cite{giambagli2021machine}. Under the spectral viewpoint, the optimization process is carried out in reciprocal space, the target of the learning being represented by the eigenvalues and eigenvectors of the square adjacency matrices that implement the information transfer in between consecutive layers of the neural network arrangement. Among other things, this allows to unequivocally identify a limited subset of parameters deemed relevant and to isolate a minimal network structure that mirrors the true complexity of the the problem under inspection \cite{, SpectralPrune, GiambagliNeurips2023}.

To restate briefly, neural networks aim at capturing  and further generalizing the patterns found in the subset of data seen during learning stages, so that they can perform similarly on data of the same typology that were not supplied during the training phase.  Overfitting occurs when the model fits too closely to the training dataset instead, thus losing the ability to generalize beyond the limited horizon of the scanned instances. The solution to prevent this unsuited effect, which dramatically limits the predictive power of the trained network, is to partition the data into three sets: a training set, a validation set, and a test set. Typically the training set accounts for 70 / 80 \% of the full data load, and the test and validation are equally split among the rest. The idea is to train the neural network on the set of training data and to evaluate its performance on the validation set in order to find the optimal hyperparameters (external variables that are fed to the model training) and related stopping criterion (number of epochs, or complete pass of the training dataset through the algorithm). In essence, the validation set is used as an independent dataset to gauge the model's performance across different model and hyperparameter choices. Overfitting can be spotted  (and thus consequently cured) by looking at validation metrics, e.g. loss or accuracy measured against the validation set. Usually, the validation metric stops improving after a certain number of epochs and begins to decrease afterward, while the training metric (namely the metric computed on the training set) keep on improving because the model seeks to (over)adjust to the limited (though large) set of instances that are handled during optimization process. Identifying the condition (as e.g. the maximum number of allowed iterations of the recursive optimization process) beyond which the performance start deteriorating due to overfitting is crucial to fine tune the network that should operate on the test set. This latter is a separate subset of the data withheld during the training phase. It is an unbiased benchmark that can be used to evaluate the model's performance after training and against a real-world data that were not examined during training stages.
 
In the following we will review two application fo neural network based techniques to the Earth Sciences.


\section{Application to volcanology}\label{sec3}

Understanding the dynamics of active volcanic systems is crucial to volcanic hazard assessment and can set the basis to forecast volcanic eruptions. This includes the geochemical and rheological characteristics of primary and evolved magmas, crustal storage conditions (P-T), ascent dynamics and interaction with crustal rocks, timescales of pre-eruptive magmatic processes, along with geophysical surveys \cite{scaillet2008upward,kahl2015constraints,saccorotti2015situ, petrone2016pre, cashman2017vertically, lin2018seismic, magee2018magma, mohamed2022geometry,rasmussen2022magmatic}.
In the last decades the understanding of magma storage, how it migrates and feeds volcanoes has changed from a melt-dominated magma chamber to more complex networks of lenses of melts, crystal mushes, and exsolved volatiles that extend throughout the crust up to the uppermost mantle \cite{cashman2017vertically}. This network plays a pivotal role in the eruption dynamics triggered by influx of fresh, hot, and deeper magmas into shallower reservoirs through prolonged magma segregation, stoping, and crystallization. Having this scenario in mind, thermobarometry is fundamental to enable grasping the anatomy of the plumbing system of  volcanoes, and provide key information on the depth of magmatic reservoirs and related migration. This proves of paramount importance for volcanic hazard assessment. Among minerals employed to assess magma storage conditions, clinopyroxene is by far the best available option, as it is widespread in mafic to intermediate and even evolved magmas \cite{putirka2008thermometers,masotta2013clinopyroxene,neave2017new}.

In \cite{chicchi2023frontiers} we have developed and thoroughly tested
GAIA (Geo Artificial Intelligence thermobArometry), a novel geothermobarometer based on clinopyroxene-only compositions which proves able to assess P-T conditions of magma storage and ascent within the crust. By exploiting a Feedforward Neural Network (FNN), GAIA returns predictions that are more robust than  thermodynamically-based clinopyroxene-liquid geothermobarometers \cite{putirka2008thermometers,masotta2013clinopyroxene, neave2017new, neave2019clinopyroxene} and other machine learning-based geothermobarometers developed in the last years \cite{petrelli2020machine,li2022machine,higgins2022machine,jorgenson2022machine}.

In the setting of the model we operated with LEPR dataset \cite{hirschmann2008library}, supplemented with a recent compilation of experimental petrology data. The dataset is made of clinopyroxene-liquid pairs, spanning a range of about 40 years, for about 6800 independent entries. The dataset has been manipulated with the application of few selected filters. In particular, data with pressure entries in between 1 bar and 10 kbar (6118 data) have been solely considered. When  it comes to temperature, we limited the analysis to datapoints with values in the range$<$1500°C (6689 data). Further, melt composition was restricted to SiO2 $>$ 35 wt.\% (6190 data) and MgO $<$20wt.\% (6606 data) to access a wide spectrum of melts, from mafic to felsic compositions. Clinopyroxenes were filtered on the basis of multiple parameters to account for the quality of the chemical analysis  and include mainly quadrilater compositions and a few Ca-Na pyroxenes. The details on the applied filters can be found in \cite{chicchi2023frontiers}. Eventually, the combination of all applied filters restricted the database to be used to 5594 clinopyroxene-melt pairs and 5599 clinopyroxenes, which were fed to the FNN for further scrutiny. Notice that the FNN is here used in regression mode, as the target of the analysis is the prediction of just two scalars, one for the temperature and the other for the pressure, over a continuum (though finite and reasonably constrained) range of variability.

The available dataset was initially split into two distinct portions: the data points belonging to one group define the so-called test set, while the elements of the other are used to train the neural network, following the lines of the above reasoning (we will return later on discussing the role of the validation set). 

Two different fractions of the dataset were kept aside as test set for temperature and pressure regression. At variance wit temperature, pressure is in fact heterogeneously distributed across the explored range (1 bar $<$ P $<$ 10 kbar). For this reason, and to maximize the chance that sparsely populated pressure ranges are adequately represented in the training set, we chose to include in the test set only 5\% of the total dataset. In contrast, we have opted for using a larger test set (20\% of the total dataset) to tune the FNN for temperature predictions. This is mainly because, the temperature dataset is evenly distributed across the considered temperature range ($<$1500°C). Data were first normalized to belong to the interval $[0,1]$. This was achieved by dividing each individual instance by the corresponding maximum, that is the largest value displayed by the homologous quantities defining the training set. A bootstrap procedure was then implemented, as detailed hereafter. The global training set was split into two parts, the actual training set and the validation set. The training set contains a number of items equal to 80\% of the total. The instances that populate the set are sampled with replacement. The elements that have not been selected (at least 20\% of the total) define the validation set. This operation is repeated $M$ times and each configuration employed to train a selected model. After training has been completed, we are hence left with $M$ distinct feedforward models that can be challenged against data. The accuracy of the predictions can be assessed versus the validation set (for each of the trained model and to prevent overfitting) and, more importantly, against the test set, that has been protected from further scrutiny at the beginning of the procedure. Dealing with $M$ distinct trained feedforward networks allows for a careful assessment of the prediction of the uncertainty (1$\sigma$) associated to each analyzed instance. 

Data were processed via a deep neural network with a feedforward architecture \cite{sebe2005machine,graves2013speech, sutton2018reinforcement, grigorescu2019survey}, of the type illustrated in the preceding section. Two separate networks were used for temperature and pressure regressions. The architecture was chosen after careful evaluation of the performance of different candidate models against validation set. In particular, a network consisting of three hidden layers of 1000 nodes each was used for temperature regression, while a network made of three hidden layers of 100 nodes each was employed to forecast the associated pressure. In both applications, a batch normalization (a dedicated normalization of the outputs in each layer to enhance training stability and convergence) is performed and a dropout layer (a regularization technique that randomly deactivates neurons during training, so as to prevent overfitting and promoting generalization) with rate set to 0.1 was inserted, after the first hidden layer.  The last layer is linear with no activation function. The model is hence not forced to yield positive forecasts, meaning that negative predictions for the estimated quantities are not a priori excluded (the fact that the estimated values are found to populate the positive semi-axis, except for a few outliers for the case of the pressure, can be rationalized as an a posteriori indication to attest the correctness of the implemented procedure). Each training iteration consisted in 500 epochs with a batch size (the number of samples that will be propagated through the network) fixed to 50. The Adam algorithm - an adaptive learning rate algorithm designed to improve the speeds of neural networks training and reach quickly the asymptotic convergence \cite{kingma2017adam} - has been used as optimizer, with a learning rate of $5 \times 10^{-5}$. 

\begin{figure}[ht]
	\centering
	\includegraphics[width = 1\textwidth]{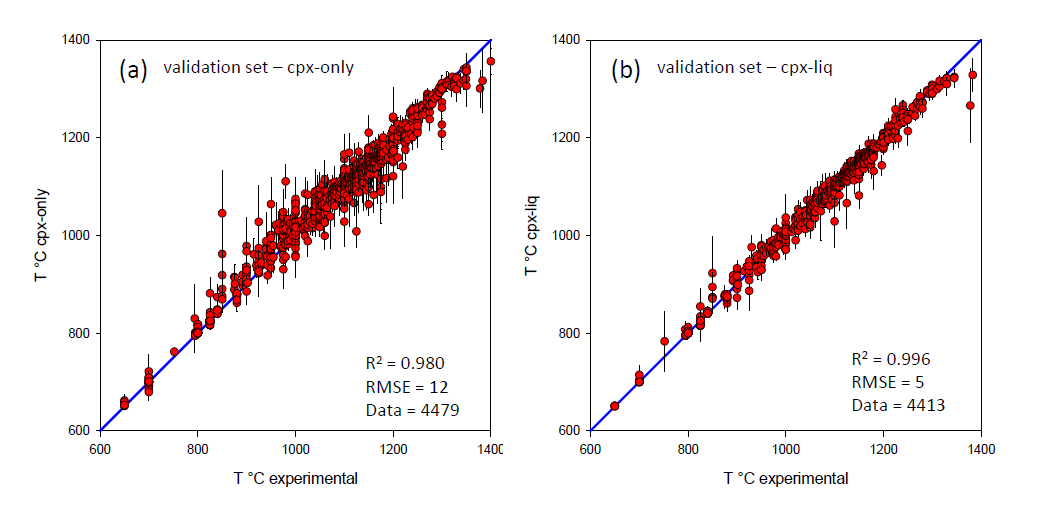}
	\caption{\footnotesize Predicted temperature of the validation set vs. experimental temperature using the 	Feedforward Neural Network method. (a) Clinopyroxene-only geothermometer, and (b) linopyroxene-liquid geothermometer. Each data point represents the average prediction and associated uncertainty (1$\sigma$) computed after $M$ independent replicas trained with the bootstrap procedure (see text).Regression statistics [R2 score, Root-Mean-Square Error (RMSE), and number of data (n)] are also reported in the displayed insets. Images taken from \cite{chicchi2023frontiers}.}
\label{vulcani_fig1}
\end{figure}

\begin{figure}[ht]
	\centering
	\includegraphics[width = 1\textwidth]{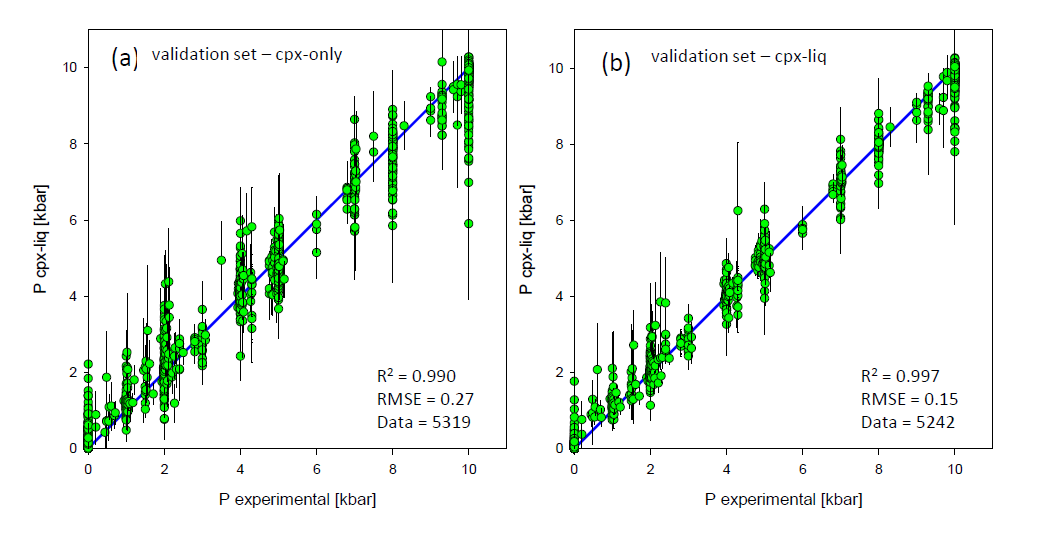}
	\caption{\footnotesize Predicted pressure of the validation dataset vs. experimental pressure using the Feedforward Neural Network method. (a) Clinopyroxene-only geobarometer, and (b) clinopyroxene-liquid geobarometer. Each data point represents the average prediction and associated uncertainty (1$\sigma$) 	computed after M independent replicas trained with the bootstrap procedure (see text). Regression 	statistics [R2 score, Root-Mean-Square Error (RMSE), and number of data (n)] is reported in the two insets. Images taken from \cite{chicchi2023frontiers}.
}
\label{vulcani_fig2}
\end{figure}

We empirically found that, a larger network is required for the temperature to be adequately predicted. We believe that this is somehow related to the distribution of the input temperatures which cover - almost uniformly - a bound, though extended domain, across the (positive) real axis. Conversely, pressures are adequately predicted by means of a significantly smaller network. This is probably due to the fact that, within the analyzed dataset, pressures tend to cluster around a few representative values. Results of the analysis (for the validation test) are displayed in Figs \ref{vulcani_fig1} and \ref{vulcani_fig2}. They prove stable for networks larger than those employed in our analysis, at fixed architecture. 

\begin{figure}[ht]
	\centering
	\includegraphics[width = 1\textwidth]{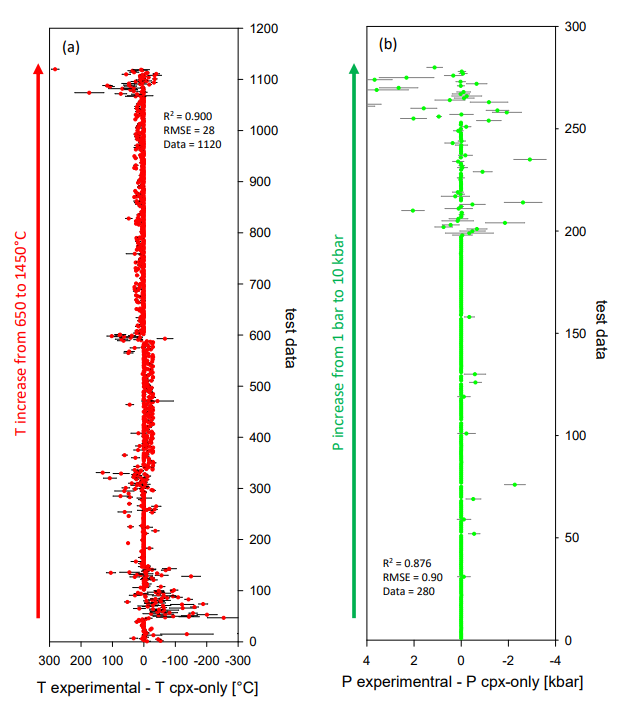}
	\caption{\footnotesize  Difference between experimental  predicted intensive parameters of the test set using the Feedforward Neural Network method applied to clinopyroxene-only. (a) temperature, test set 20\% of the total available dataset, (b) pressure, test set 5\% of the total available dataset. Each data point represents the average difference and associated uncertainty (1$\sigma$) computed after $M$ independent replicas trained with the bootstrap procedure (see text). Regression statistics [R2 score, Root-Mean-Square Error (RMSE), and number of data (n)] are also reported in the annexed insets. Data points have been displayed from the lowest T and P values (bottom of the y-axis) to the highest T and P values (top of the y-axis) to help visualization.  This is particularly	important in the case of pressure because of its uneven distribution (localized at 1 bar, 2 kbar, 4 kbar, 8 kbar, 10	kbar). Images taken from \cite{chicchi2023frontiers}.
}
\label{vulcani_figS3}
\end{figure}

The accuracy and the computed Root-Mean-Square Error (RMSE) suggests that the FNN method to estimate P and T can be reliably used against clinopyroxene data only. The uncertainties on P-T estimates are slightly worse than those of obtained by employing clinopyroxene-liquid pairs, but working under the former scenario that solely accounts for clinopyroxene we avoid facing the longstanding dilemma of the crystal-liquid equilibrium. Motivated by this success, we have then turned to checking the performance of the method against data that were kept aside from the training procedure (the test set). The P-T predictions (based on just clinopyroxene data) are reported in the two panels of Fig. \ref{vulcani_figS3} in terms of the difference between experimental and estimated P-T,and confirm the adequacy of the proposed approach.

The trained model has been also applied to  assess the P-T conditions of magma storage and ascent within the crust to five active Italian volcanoes, namely: Etna and Stromboli (currently active), Campi Flegrei and Vulcano-Vulcanello (in volcanic unrest), and Somma-Vesuvio (quiescent). Their clinopyroxene compositions were downloaded from the GEOROC Data Repository (https://georoc.eu/georoc), and the obtained results are reported in \cite{chicchi2023frontiers}.

To make the model easily accessible, we have developed a web free and open-source app. The app can be reached at this link: https://gaia-geothermobarometry-gaia-home-6ol8kg.streamlit.app/. The app will begin by calculating the components of the clinopyroxenes which will be fed to the neural networks. At the end of the process, an output file will be provided with the corresponding estimates for pressure and temperature.

\section{Application to superconductors}\label{sec4}

Finding new superconductors proves a complex task, which heavily relies on operators' experience and intuition. By sampling from a large gallery of candidate materials, it was indeed shown that just a limited fraction of the total, in the range of 3 \% \cite{Hosono2015}, shows a superconducting phase. Such a modest score severely limit our actual ability to effectively chase for novel superconductors via conventional approaches. Machine learning (ML) technologies holds the potential to fill this gap by providing an innovative angle to tackle the problem of identifying putative superconducting materials.  

Preliminary attempts to apply ML technology to a large superconducting database have been reported in \cite{Stanev2018, Ho95, Konno2021}. 
Particularly interesting is the work \cite{Konno2021} where
the regression problem (anticipating the critical temperature of the examined material) was brought back to a standard image processing. More specifically, the stoichiometries of the inspected materials  are entered into the two-dimensional periodic table, which is then supplied to the convolutional layers to resolve the relative positions of the constituting elements. A possible drawback of this procedure is that the outcome of the analysis may sense the specific arrangement of the periodic table, thus materializing in unsuited bias.

Starting from these premises, in \cite{Pereti} we have proposed a 
different approach to the supervised classification and regression of superconductive materials which exploits the celebrated Deep Sets technology \cite{Komiske2019}. Deep Sets are a family of deep learning algorithms specifically designed to operate on \emph{sets} made of  finite (distinct) members or elements. A valid function defined on a set should be invariant under permutations of the elements of the set itself. Stated differently, each permutation that reshuffles a pre-existing order, should return the very same functional output and thus, the produced output should be insensitive to the imposed permutation. In concrete terms, the Deep Set takes as an input a set (of variable dimension) that reflects the chemical composition of the examined compound. For the specific case, each element is characterised by a list of $22$ instances, and specifically: the atomic number,  the atomic volume, the block in periodic table, the density, the polarizability, the electron affinity, the evaporation heat, the fusion heat, the group in periodic table, the lattice constant, the lattice structure symbol, the melting temperature, the period in the periodic table, the specific heat at 293.15 K ($20^{o}$C), the thermal conductivity at 293.15 K ($20^{o}$C), the Van der Waals radius, the covalent radius, the Pauling's scale of electronegativity, the atomic weight, the atomic radius, the ionization energies (in eV), the valence. The stoichiometric integer of the selected atom is also supplied, as an additional information at the entry level. Each input element of the set is transformed by a mapping function,  encoded in a deep neural network (of the type discussed in above) to be trained. The resulting elements are summed up (pooling operation) and the obtained output further processed by another non-linear filter, which takes the form of a additional, fully trainable, multi-layered neural network. In \cite{Pereti} the interested reader can find a schematic layout of the employed Deep Set architecture. 

To train the DeepSet we made use of the data reported in the SuperCon database, which consists of Oxide and Metallic compounds (inorganic materials containing metals, alloys, cuprate high-temperature superconductors, etc.) and Organic compounds (organic superconductors). Removing incomplete entries and multiple records yielded a final collection of 16395 independent elements. 

For the regression task (i.e. determining the critical temperature $T_c$), the SuperCon database was split into training (80 \% of the total) and test  (the residual 20 \%) sets. This partition is random and thus the experiments (training and consequent predictions upon evaluation on the residual test) can be repeated ($50$ times in our analysis) to yield average prediction with associated errors. 

For classification (i.e. determining if a given compound is superconducting or not),  we adopted the so called "Garbage in" technique as introduced \cite{Konno2021}, which in turn allows one to access a larger set of materials that are presumably non superconductors. In short, the idea is to complement the Supercon database with the inclusion of an additional pool of elements drawn from the Crystallography Open Database (COD). Since superconductors are just a few percent of the total, these latter materials are labelled as non superconductors. The inclusion of superconductors in the reservoir of non superconducting materials are, in all probability, small and it can be reasonably postulated that the benefits of enlarging the set of scrutinized materials, overcome the disadvantages produced by the "garbage" (the erroneous entries) that was put deliberately in. 

 Another dataset that we employed is that assembled by 
 Hosono et al in \cite{Hosono2015}. This latter is composed by a relatively short list of materials that experts with a background in solid state chemistry have identified as bearing the potential traits of superconductors. After experimental scrutiny, a limited subset of the total was indeed confirmed to display a superconducting phase. More precisely, the database is made of 207 materials, 39 superconductors and 168 non superconductors. It should be remarked that the Hosono database allows a rather stringent test of the performance of Deep Set classificator, since materials have been pre-selected, by recognized experts in the field, as possible superconductors.

\begin{figure}
\centering
\includegraphics[scale=0.9]{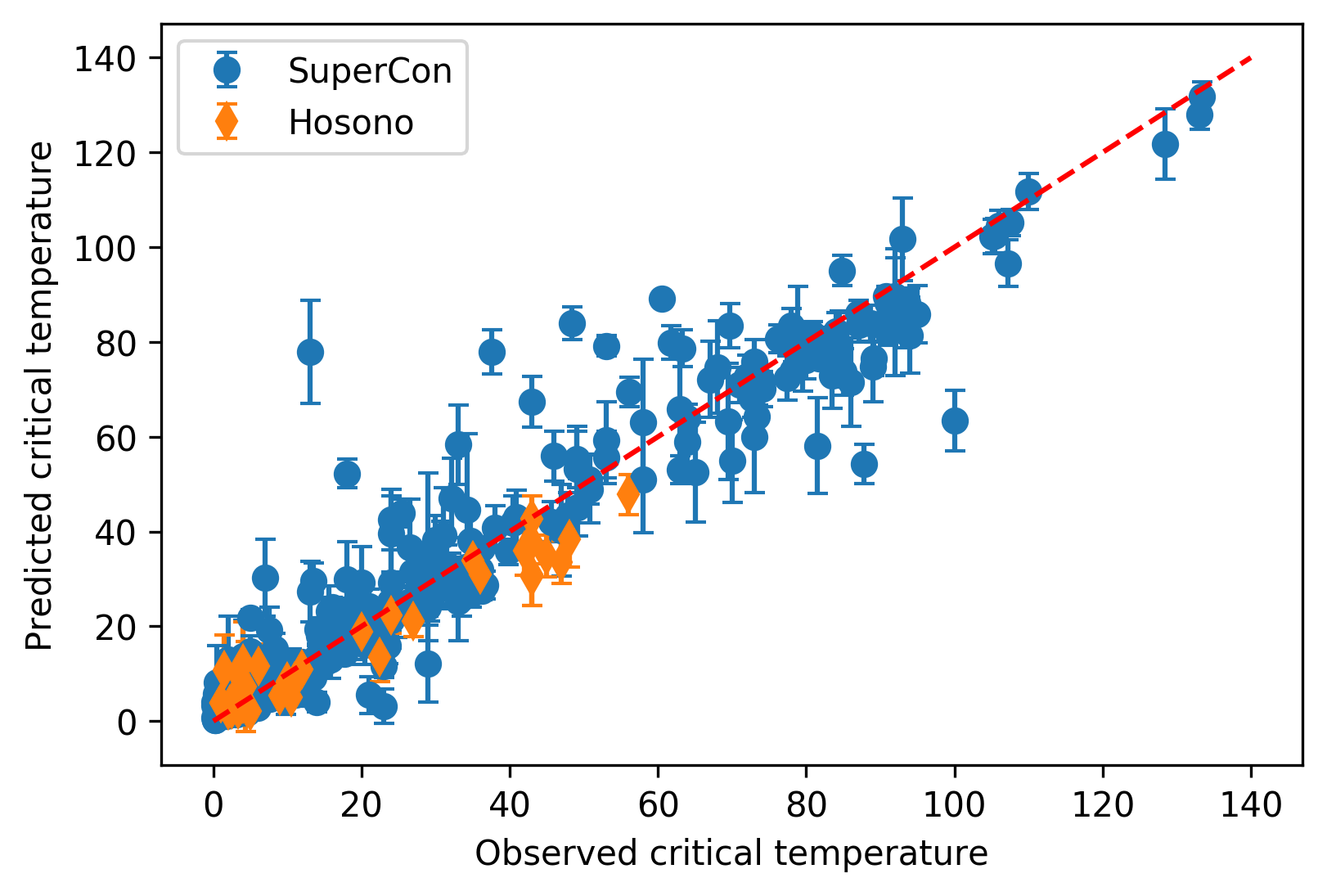} 
\caption{Predicted vs. measured critical temperature, $T_c$ (K). The blue circles refer to materials from the SuperCon database. A fraction of 20 \% of the total is 
randomly selected as part of the test set, while the remaining 80 \% is used for training. The operation is repeated 50 times and the collected temperatures from each independent run stored for further processing. The displayed data refer to the 328 materials selected to be part of the test set at least 10 times, out of  50 trials. The orange diamonds refer to the 39 superconducting materials (all tested 50 times by independent DeepSet realizations) that are present in the  Hosono database. This latter contains a total of 207 (superconducting and non-superconducting) compounds.  Symbols stand for the average temperature, while the error bars are the computed standard deviation. Image taken from \cite{Pereti}.}
\label{figure2}
\end{figure}

In Fig. \ref{figure2}, the predicted $T_c$ is reported against the corresponding values, as measured experimentally. Blue circles refer to materials of the SuperCon dataset \cite{Konno2021} that were not employed for the training process. The computed errors is calculated by combining data from from 50 independent realisations of the trained device (only materials that have been selected at least 10 times are considered). As it can be appreciated by visual inspection, data align along the diagonal (dashed line) of the first quadrant thus confirming the adequacy of the proposed procedure. The  root mean squared error  is equal to 9.5 K, while $r^2=0.92$. 

The diamonds displayed in orange refer to the elements of the Hosono database \cite{Hosono2015, Konno2021}. In this case, the estimated root mean squared error is found to be 7 K ($r^2=0.84$). In summary, and based on the results displayed in Fig. \ref{figure2},  the trained Deep Set proved capable to satisfactorily anticipate the critical temperature of a superconducting material, starting from its description in terms of constitutive elements.

Let us now turn to the classification task. The Deep Set was trained on a sub-portion of the SuperCon database, potentiated with the addition of data from the Crystallography Open Database (COD), as follows the  "garbage in" procedure. The trained model is then used against the test-sets of interest, as e.g. the Hosono database. To improve on the system performance we also made used of an apt majority rule, applied to the set of recorded outcomes (further details can be found in \cite{Pereti}).
By operating under this scheme, we successfully identified 29 out of 39 superconductors of the Hosono database, with 19 false positive, a  particularly satisfactory result given the fact that the Hosono database contains pre-filtered samples, and thus all materials are in principle good superconducting candidates.

Finally, we used the trained Deep Set against the updated list (September 2021) of minerals accepted by the International Mineralogical Association \cite{IMA}. 
The list of minerals classified as superconductors is made available in the Supplementary Information annexed to \cite{Pereti}. Interestingly, about 44 \% of the minerals selected by the trained network were indeed already checked as superconductors.
We then focused on a subset of (three) candidate materials, that have not yet been characterized - or poorly characterized \cite{micheneriteEXP} - in the literature, with reference to their potential superconducting behaviour. These are the synthetic analogues of  Pd$_3$HgTe$_3$ (the analogue of the mineral temagamite), PdBiTe (the analogue of the mineral michenerite), and Pd$_2$NiTe$_2$ (the analogue of the mineral monchetundraite). As concerns the critical temperatures, we predicted $T_c$ of 1.8(1.8) K, 1.6(0.8) K and 1.18(0.7) K, respectively. We employed magnetic ac susceptometry to investigate the synthesized analogues of the three minerals expected to exhibit superconductivity. The two examined samples of temagamite  did not reveal any sizeable diamagnetic susceptibility, while a clear onset of diamagnetic shielding was clearly detectable  at \textit{T}=2.10 K for PdBiTe and, even more pronounced, at \textit{T}=1.06 K for Pd$_2$NiTe$_{2}$ (monchetundraite). For a detailed account of the experimental characterization of the above material the reader can refer to \cite{Pereti}. Superconductivity has been thus confirmed for the synthetic analogue of michenerite, PdBiTe,
and observed for the first time in monchetundraite, Pd$_2$NiTe$_{2}$, at critical temperatures in excellent agreement with that predicted by our trained algorithm. The monchetundraite can be hence regarded, to the best of our knowledge, as the first certified superconducting material to be effectively discovered by artificial intelligence methodologies.

\section{Conclusion}\label{sec5}

Neural networks of different conceptions are gaining progressive  importance as innovative  methodological research tools. Motivated by this observation, we have here provided a simple description of the basic functioning of a feedforward neural network. This a collection of neurons, the elementary computing units of the network, arranged in different stacks or layers, sequentially ordered from the input (the location where the instance to be analyzed is provided) to the output (where decision is eventually made). In the first part of the paper, we synthetically described the actual functioning of such a computational devise, with some reference to the algorithmic aspects of the training phase. Then, we moved forward to revise two recent applications to 
Earth and Materials Science. We have in particular discussed a novel Machine Learning based solution to estimate P-T conditions of magma  storage and migration within the crust \cite{chicchi2023frontiers}. The predictive adequacy of the trained model, as testified by direct evaluation of the associated performance, suggests that neural networks could materialize in a novel class of reliable geothermobarometers. As a second application, we have reviewed the work in \cite{Pereti} which aimed at the building an automated algorithm for supervised classification and regression of superconductive (SC) materials. In this case the idea is to anticipate the possible superconducting nature (and quantify its associated critical temperature) of a given sample by providing to the trained algorithm its chemical composition. This computed assisted classification applied to the whole list of  minerals from the International Mineralogical Association, allowed us to discover the first certified superconducting material identified by artificial intelligence methodologies. Summing up, feedforward neural networks, and their diverse branches, provide a powerful tools which, when properly mastered and critically supervised,  can help boosting the research in different domains, including those that pertain to the vast realm of Earth Science applications.









\bibliography{sn-bibliography}

\end{document}